\documentclass[aps,pra,showpacs,superscriptaddress,preprint]{revtex4}
\usepackage{amsmath}
\usepackage{epsf}
\usepackage{epsfig}

\begin{document}
\title{{\bf An improved approximation to }$l${\bf -wave bound states of the
Manning-Rosen potential by Nikiforov-Uvarov method }}
\author{Sameer M. Ikhdair\thanks{%
sameer@neu.edu.tr} and \ Ramazan Sever\thanks{%
sever@metu.edu.tr}}
\address{$^{\ast }$Department of Physics, Near East University, Nicosia, Cyprus,
Turkey. \\
$^{\dagger }$Department of Physics, Middle East Technical University, 06531
Ankara, Turkey.}
\date{\today}

\begin{abstract}
A new approximation scheme to the centrifugal term is proposed to obtain the
$l\neq 0$ solutions of the Schr\"{o}dinger equation with the Manning-Rosen
potential. We also find the corresponding normalized wave functions in terms
of the Jacobi polynomials. To show the accuracy of the new approximation
scheme, we calculate the energy eigenvalues numerically for arbitrary
quantum numbers $n$ and $l$ with two different values of the potential
parameter $\alpha .$ The bound state energies of various states for a few $%
HCl,$ $CH,$ $LiH$ and $CO$ diatomic molecules are also calculated. The
numerical results are in good agreement with those obtained by using program
based on a numerical integration procedure. Our solution can be also reduced
to the $s$-wave ($l=0$) case and to the Hulth\'{e}n potential case.


PACS NUMBER(S): 03.65.-w; 02.30.Gp; 03.65.Ge; 34.20.Cf
\end{abstract}
\maketitle
\bigskip

\section{Introduction}

\noindent The exact analytic solutions of the wave equations
(nonrelativistic and relativistic) are only possible for certain potentials
of physical interest under consideration since they contain all the
necessary information on the quantum system. It is well known that the exact
solutions of these wave equations are only possible in a few simple cases
such as the Coulomb, the harmonic oscillator, pseudoharmonic potentials and
others [1-5]. Recently, the analytic exact solutions of the wave equation
with some exponential-type potentials are impossible for $l\neq 0$ states.
Approximation methods have to be used to deal with the centrifugal term like
the Pekeris approximation [6-8] and the approximated scheme suggested by
Greene and Aldrich [9]. Some of these exponential-type potentials include
the Morse potential [10], the Hulth\'{e}n potential [11], the
P\"{o}schl-Teller [12], the Woods-Saxon potential [13], the Kratzer-type and
pseudoharmonic potentials [14], the Rosen-Morse-type potentials [15], the
Manning-Rosen potential [15-19] and other multiparameter exponential-type
potentials [20,21].

The Manning-Rosen (M-R) potential has been one of most useful and convenient
models to study the energy eigenvalues of diatomic molecules [16]. As an
empirical potential, the M-R potential gives an excellent description of the
interaction between the two atoms in a diatomic molecule and also it is very
reasonable to describing the interactions close to the surface. The short
range M-R potential is defined by [15-19]

\begin{equation}
V(r)=-\frac{A\hbar ^{2}}{2\mu b^{2}}\frac{e^{-r/b}}{1-e^{-r/b}}+\frac{\alpha
(\alpha -1)\hbar ^{2}}{2\mu b^{2}}\left( \frac{e^{-r/b}}{1-e^{-r/b}}\right)
^{2},
\end{equation}
where $A$ and $\alpha $ are two-dimensionless parameters [22] but the
screening parameter $b$ has dimension of length which has a potential range $%
1/b.$ The potential (1) may be further put in the following simple form

\begin{equation}
V(r)=-\frac{Ce^{-r/b}+De^{-2r/b}}{\left( 1-e^{-r/b}\right) ^{2}},\text{ }C=A,%
\text{ }D=-A-\alpha \text{(}\alpha -1)\text{,}
\end{equation}
which is usually used for the description of diatomic molecular vibrations
[23,24]. It is also used in several branches of physics for their bound
states and scattering properties. The potential in (1) remains invariant by
mapping $\alpha \rightarrow 1-\alpha $ and has a relative minimum value $%
V(r_{0})=-\frac{A^{2}}{4\kappa b^{2}\alpha (\alpha -1)}$ at $r_{0}=b\ln %
\left[ 1+\frac{2\alpha (\alpha -1)}{A}\right] $ for $\alpha >0$ to be
obtained from the first derivative $\left. \frac{dV}{dr}\right|
_{r=r_{0}}=0. $ The second derivative which determines the force constants
at $r=r_{0}$ is given by

\begin{equation}
\left. \frac{d^{2}V}{dr^{2}}\right| _{r=r_{0}}=\frac{A^{2}\left[ A+2\alpha
(\alpha -1)\right] ^{2}}{8b^{4}\alpha ^{3}(\alpha -1)^{3}}.
\end{equation}
It is known that for this potential the Schr\"{o}dinger equation (SE) can be
solved for the $s$-wave, angular momentum quantum number $l=0.$ However, in
general solution, it is needed to include some approximations if one wants
to obtain analytical or semianalytical solutions to the SE. Also, it is
often necessary to determine the $l$-wave ($l\neq 0$ states), so an analytic
procedure would be advantageous [25-27]. Hence, in the previous papers,
several approximations have been developed to find better analytical
formulas for the M-R potential. For instance, in the $l=0$ case, the
bound-state energy eigenvalues for the M-R potential have already been
calculated by using the path-integral approach [17] and function analysis
method [18]. For the $l\neq 0$ case, the potential can not be solved exactly
without approximation. Recently, Qiang and Dong [19] approximated the
centrifugal term
\[
\frac{1}{r^{2}}\approx \frac{1}{b^{2}}\left[ \frac{e^{-r/b}}{1-e^{-r/b}}%
+\left( \frac{e^{-r/b}}{1-e^{-r/b}}\right) ^{2}\right] =\frac{1}{b^{2}}\frac{%
e^{-r/b}}{\left( 1-e^{-r/b}\right) ^{2}}
\]
and studied $l$-wave bound state solutions of the SE with M-R potential. Wei
{\it et al.} [25] investigated the scattering state solutions of the SE with
M-R potential using the approximation [9,11,19]. Ikhdair and Sever
[11,26,27] applied the above approximation and obtained the $l$-wave
solutions of SE with the M-R potential in three-dimensions and $D$%
-dimensions and also with the Hulth\'{e}n potential using Nikiforov and
Uvarov (N-U) method. This approximations provide good results which are in
agreement with the numerical integration method by Lucha and Sch\"{o}berl
[28] for short-range potential (large $b$ and small $l$) but not for
long-range potential (small $b$ and large $l$)$.$

Our aim is to improve the accuracy of our previous approximation [26,27], so
that we propose and apply a new approximation scheme for the centrifugal
term to get our results in high agreement with Ref. [28]. Thus, with this
new approximation scheme, we calculate the $l\neq 0$ energy levels and
wavefunctions of the M-R potential using the Nikiforov and Uvarov (N-U)
method which has shown its power in calculating the exact energy levels for
some solvable quantum systems. For this, the results are in better agreement
with those obtained by means of numerical integration method [28]. As an
illustration, the method is applied to find the energy levels of the $HCl,$ $%
LiH$, $CH$ and $CO$ diatomic molecules.

The paper is organized as follows: In Section II we breifly present the
Nikiforov-Uvarov (N-U) method. In Section III, we present the new proposed
approximation scheme and apply it to calculate the $l$-wave bound state
eigensolutions of the SE with M-R potential by the N-U method. In Section
IV, we present our numerical cresults for various diatomic molecules.
Section V, is devoted to for two special cases, namely, $s$-wave ($l=0)$ and
the Hulth\'{e}n potential. Finally, we make a few concluding remarks in
Section VI.

\section{\noindent The Nikiforov and Uvarov method}

The N-U method is based on solving the second-order linear differential
equation by reducing it to a generalized equation of hypergeometric type
[29]. In this method after employing an appropriate coordinate
transformation $z=z(r),$ the Schr\"{o}dinger equation can be written in the
following form:
\begin{equation}
\psi _{n}^{\prime \prime }(z)+\frac{\widetilde{\tau }(z)}{\sigma (z)}\psi
_{n}^{\prime }(z)+\frac{\widetilde{\sigma }(z)}{\sigma ^{2}(z)}\psi
_{n}(z)=0,
\end{equation}
where $\sigma (z)$ and $\widetilde{\sigma }(z)$ are the polynomials with at
most of second-degree, and $\widetilde{\tau }(s)$ is a first-degree
polynomial. The special orthogonal polynomials [29] reduce Eq. (4) to a
simple form by employing $\psi _{n}(z)=\phi _{n}(z)y_{n}(z),$ and choosing
an appropriate function $\phi _{n}(z).$ Consequently, Eq. (4) can be reduced
into an equation of the following hypergeometric type:

\begin{equation}
\sigma (z)y_{n}^{\prime \prime }(z)+\tau (z)y_{n}^{\prime }(z)+\lambda
y_{n}(z)=0,
\end{equation}
where $\tau (z)=\widetilde{\tau }(z)+2\pi (z)$ (its derivative must be
negative) and $\lambda $ is a constant given in the form

\begin{equation}
\lambda =\lambda _{n}=-n\tau ^{\prime }(z)-\frac{n\left( n-1\right) }{2}%
\sigma ^{\prime \prime }(z),\text{\ \ \ }n=0,1,2,...
\end{equation}
It is worthwhile to note that $\lambda $ or $\lambda _{n}$ are obtained from
a particular solution of the form $y(z)=y_{n}(z)$ which is a polynomial of
degree $n.$ Further, $\ y_{n}(z)$ is the hypergeometric-type function whose
polynomial solutions are given by Rodrigues relation

\begin{equation}
y_{n}(z)=\frac{B_{n}}{\rho (z)}\frac{d^{n}}{dz^{n}}\left[ \sigma ^{n}(z)\rho
(z)\right] ,
\end{equation}
where $B_{n}$ is the normalization constant and the weight function $\rho
(z) $ must satisfy the condition [29]

\begin{equation}
\frac{d}{dz}w(z)=\frac{\tau (z)}{\sigma (z)}w(z),\text{ }w(z)=\sigma (z)\rho
(z).
\end{equation}
In order to determine the weight function given in Eq. (8), we must obtain
the following polynomial:

\begin{equation}
\pi (z)=\frac{\sigma ^{\prime }(z)-\widetilde{\tau }(z)}{2}\pm \sqrt{\left(
\frac{\sigma ^{\prime }(z)-\widetilde{\tau }(z)}{2}\right) ^{2}-\widetilde{%
\sigma }(z)+k\sigma (z)}.
\end{equation}
In principle, the expression under the square root sign in Eq. (9) can be
arranged as the square of a polynomial. This is possible only if its
discriminant is zero. In this case, an equation for $k$ is obtained. After
solving this equation, the obtained values of $k$ are included in the N-U
method and here there is a relationship between $\lambda $ and $k$ by $%
k=\lambda -\pi ^{\prime }(z).$ After this point an appropriate $\phi _{n}(z)$
can be extracted from the condition

\begin{equation}
\frac{\phi ^{\prime }(z)}{\phi (z)}=\frac{\pi (z)}{\sigma (z)}.
\end{equation}

\section{Analytical Solutions}

\subsection{An Impoved Approximation Scheme}

The approximation is based on the expansion of the centrifugal term in a
series of exponentials depending on the intermolecular distance $r$ and
keeping terms up to second order. Therefore, instead of using the
approximation in [9,11,19], we use this choice of approximation:
\[
\frac{1}{r^{2}}\approx \frac{1}{r_{0}{}^{2}}\left[
c_{0}+c_{1}v(r)+c_{2}v^{2}(r)\right] ,\text{ }v(r)=\frac{e^{-r/b}}{1-e^{-r/b}%
}
\]
\begin{equation}
\frac{1}{r^{2}}\approx \frac{1}{r_{0}{}^{2}}\left[ c_{0}+c_{1}\frac{1}{%
e^{r/b}-1}+c_{2}\frac{1}{\left( e^{r/b}-1\right) ^{2}}\right] ,
\end{equation}
which has a similar form of the M-R potential. Changing the coordinate to $x$
by using $x=(r-r_{0})/r_{0},$ one obtains

\begin{equation}
\left( 1+x\right) ^{-2}=\left[ c_{0}+\frac{c_{1}}{e^{\gamma (1+x)}-1}+\frac{%
c_{2}}{\left( e^{\gamma (1+x)}-1\right) ^{2}}\right] ,\text{ }\gamma
=r_{0}/b.
\end{equation}
and e$xpanding$ Eq. (12) around $r=r_{0}$ $(x=0),$ we obtain the following
Taylor's expansion:
\begin{equation}
\left( 1-2x+\cdots \right) =\left[ \left( c_{0}+\frac{c_{1}}{e^{\gamma }-1}+%
\frac{c_{2}}{\left( e^{\gamma }-1\right) ^{2}}\right) -\gamma \left( \frac{%
c_{1}}{e^{\gamma }-1}+\frac{c_{1}+2c_{2}}{\left( e^{\gamma }-1\right) ^{2}}+%
\frac{2c_{2}}{\left( e^{\gamma }-1\right) ^{3}}\right) x+\cdots \right] ,
\end{equation}
from which we obtain
\[
c_{0}+\frac{c_{1}}{e^{\gamma }-1}+\frac{c_{2}}{(e^{\gamma }-1)^{2}}=1,
\]
\begin{equation}
\gamma \left( \frac{c_{1}}{e^{\gamma }-1}+\frac{c_{1}+2c_{2}}{\left(
e^{\gamma }-1\right) ^{2}}+\frac{2c_{2}}{\left( e^{\gamma }-1\right) ^{3}}%
\right) =2.
\end{equation}
Taking $r_{0}=b$ ($\gamma =1$)$,$ one obtains, from Eq. (14), the following
three simple cases:

Case 1. If $c_{1}=c_{2}=1,$ then the shift, $c_{0},$ in the present
approximation is simply given by

\begin{equation}
c_{0}=1-\frac{1}{e-1}-\frac{1}{(e-1)^{2}}=0.0793264057923,
\end{equation}
where $e$ is the base of the natural logarithms, $e=2.718281828459045.$

Case 2. Without any loss of generality, we may take $c_{1}=1,$ then we can
calculate the shift $c_{0}=0.0768910877367$ and \ $c_{2}=1.007190258153.$

Case 3. If we choose $c_{2}=1,$ then we find the shift $%
c_{0}=0.0744557696812 $ and \ $c_{1}=1.0083691255228.$ Thus, for the
approximation given in case 1, we have$.$
\begin{equation}
\mathrel{\mathop{\lim }\limits_{b\rightarrow \infty }}%
\frac{1}{b^{2}}\left[ 1-\frac{1}{e-1}-\frac{1}{(e-1)^{2}}+\frac{e^{-r/b}}{%
1-e^{-r/b}}+\left( \frac{e^{-r/b}}{1-e^{-r/b}}\right) ^{2}\right] =\frac{1}{%
r^{2}}.
\end{equation}
Finally, in the case if $c_{0}=0$ and $c_{1}=c_{2}=1,$ the approximation
given in Eq. (11) is identical to the commonly used approximation in the
previous works [9,11,19,26,27].

\subsection{Bound State Solutions}

To study any quantum physical system characterized by the empirical
potential given in Eq. (1), we solve the original ${\rm SE}$ which is given
in the well known textbooks [1,2]

\begin{equation}
\left( \frac{p^{2}}{2m}+V(r)\right) \psi ({\bf r,}\theta ,\phi )=E\psi ({\bf %
r,}\theta ,\phi ),
\end{equation}
where the potential $V(r)$ is taken as the M-R form in (1). Using the
separation method with the wavefunction $\psi ({\bf r,}\theta ,\phi
)=r^{-1}R(r)Y_{lm}(\theta ,\phi ),$ we obtain the following radial
Schr\"{o}dinger eqauation as

\[
\frac{d^{2}R_{nl}(r)}{dr^{2}}+\left[ \frac{2\mu E_{nl}}{\hbar ^{2}}%
-V_{eff}(r)\right] R_{nl}(r)=0,
\]
\begin{equation}
V_{eff}(r)=\frac{1}{b^{2}}\left[ \frac{\alpha (\alpha -1)e^{-2r/b}}{\left(
1-e^{-r/b}\right) ^{2}}-\frac{Ae^{-r/b}}{1-e^{-r/b}}\right] +\frac{l(l+1)}{%
r^{2}}.
\end{equation}
Since the SE with the above M-R effective potential has no analytical
solution for $l$-waves, the approximation to the centrifugal term given by
case 1 has to be made so that the energy eigenvalues are found to be in
better agreement with those obtained by means of the numerical integration
method [28]. The other approximations will be left for future
investigations. To solve it by the N-U method, we need to recast Eq. (18)
with Eq. (16) into the form of Eq. (4) changing the variables $r\rightarrow
z $ through the mapping function $r=f(z)$ and energy transformation given by

\begin{equation}
z=e^{-r/b},\text{ }\varepsilon ^{\prime }=\sqrt{-\frac{2\mu b^{2}E_{nl}}{%
\hbar ^{2}}+\Delta E_{l}},\text{ }E_{nl}<\frac{\hbar ^{2}}{2\mu b^{2}}\Delta
E_{l},\text{ }\Delta E_{l}=l(l+1)c_{0},
\end{equation}
to obtain the following hypergeometric equation:

\[
\frac{d^{2}R(z)}{dz^{2}}+\frac{(1-z)}{z(1-z)}\frac{dR(z)}{dz}
\]

\begin{equation}
+\frac{1}{\left[ z(1-z)\right] ^{2}}\left\{ -\varepsilon ^{\prime }{}^{2}+%
\left[ A+2\varepsilon ^{\prime }{}^{2}-l(l+1)\right] z-\left[ A+\varepsilon
^{\prime }{}^{2}+\alpha (\alpha -1)\right] z^{2}\right\} R(z)=0.
\end{equation}
We notice that for bound state (real) solutions, the last equation requires
that

\begin{equation}
z=\left\{
\begin{array}{ccc}
0, & \text{when} & r\rightarrow \infty , \\
1, & \text{when} & r\rightarrow 0,
\end{array}
\right.
\end{equation}
and thus the finite radial wavefunctions $R_{nl}(z)\rightarrow 0.$ To apply
the N-U method, we compare Eq. (20) with Eq. (4) and obtain the following
values for the parameters:
\begin{equation}
\widetilde{\tau }(z)=1-z,\text{\ }\sigma (z)=z-z^{2},\text{\ }\widetilde{%
\sigma }(z)=-\left[ A+\varepsilon ^{\prime }{}^{2}+\alpha (\alpha -1)\right]
z^{2}+\left[ A+2\varepsilon ^{\prime }{}^{2}-l(l+1)\right] z-\varepsilon
^{\prime }{}^{2}.
\end{equation}
If one inserts these values of parameters into Eq. (9), with $\sigma
^{\prime }(z)=1-2z,$ the following linear function is achieved
\begin{equation}
\pi (z)=-\frac{z}{2}\pm \frac{1}{2}\sqrt{\left\{ 1+4\left[ A+\varepsilon
^{\prime }{}^{2}+\alpha (\alpha -1)\right] -k\right\} z^{2}+4\left\{ k-\left[
A+2\varepsilon ^{\prime }{}^{2}-l(l+1)\right] \right\} z+4\varepsilon
^{\prime }{}^{2}}.
\end{equation}
According to this method the expression in the square root has to be set
equal to zero, that is, $\Delta =\left\{ 1+4\left[ A+\varepsilon ^{\prime
}{}^{2}+\alpha (\alpha -1)\right] -k\right\} z^{2}+4\left\{ k-\left[
A+2\varepsilon ^{\prime }{}^{2}-l(l+1)\right] \right\} z+4\varepsilon
^{\prime }{}^{2}=0.$ Thus the constant $k$ can be determined as

\begin{equation}
k=A-l(l+1)\pm a\varepsilon ^{\prime },\text{ \ }a=\sqrt{(1-2\alpha
)^{2}+4l(l+1)}.
\end{equation}
In this regard, we can find four possible functions for $\pi (z)$ as
\begin{equation}
\pi (z)=-\frac{z}{2}\pm \left\{
\begin{array}{c}
\varepsilon -\left( \varepsilon ^{\prime }-\frac{a}{2}\right) z,\text{ \ \ \
for \ \ }k=A-l(l+1)+a\varepsilon ^{\prime }, \\
\varepsilon -\left( \varepsilon ^{\prime }+\frac{a}{2}\right) z;\text{ \ \ \
for \ \ }k=A-l(l+1)-a\varepsilon ^{\prime }.
\end{array}
\right.
\end{equation}
We must select

\begin{equation}
\text{\ }k=A-l(l+1)-a\varepsilon ^{\prime },\text{ }\pi (z)=-\frac{z}{2}%
+\varepsilon ^{\prime }-\left( \varepsilon +\frac{a}{2}\right) z,
\end{equation}
in order to obtain the polynomial, $\tau (z)=\widetilde{\tau }(z)+2\pi (z)$
having negative derivative as

\begin{equation}
\tau (z)=1+2\varepsilon ^{\prime }-\left( 2+2\varepsilon ^{\prime }+a\right)
z,\text{ }\tau ^{\prime }(z)=-(2+2\varepsilon ^{\prime }+a).
\end{equation}
We can also write the values of $\lambda =k+\pi ^{\prime }(z)$ and $\lambda
_{n}=-n\tau ^{\prime }(z)-\frac{n\left( n-1\right) }{2}\sigma ^{\prime
\prime }(z),$\ $n=0,1,2,...$ as

\begin{equation}
\lambda =A-l(l+1)-(1+a)\left[ \frac{1}{2}+\varepsilon ^{\prime }\right] ,
\end{equation}
\begin{equation}
\lambda _{n}=n(1+n+a+2\varepsilon ^{\prime }),\text{ }n=0,1,2,\cdots
\end{equation}
respectively. Additionally, using the definition of $\lambda =\lambda _{n}$
and solving the resulting equation for $\varepsilon ^{\prime },$ allows one
to obtain

\begin{equation}
\varepsilon ^{\prime }=\frac{(n+1)^{2}+l(l+1)+(2n+1)\Lambda -A}{%
2(n+1+\Lambda )},\text{ }\Lambda =\frac{-1+a}{2}.
\end{equation}
Using Eqs. (19) and (30), we obtain the discrete energy levels

\begin{equation}
E_{nl}=-\frac{\hbar ^{2}}{2\mu b^{2}}\left[ \frac{(n+1)^{2}+l(l+1)+(2n+1)%
\Lambda -A}{2(n+1+\Lambda )}\right] ^{2}+\frac{\hbar ^{2}l(l+1)c_{0}}{2\mu
b^{2}},\text{ \ }0\leq n,l<\infty ,
\end{equation}
where $n=0,1,2,\cdots $ and $l$ signify the usual radial and angular
momentum quantum numbers, respectively. It is found that $\Lambda $ remains
invariant by mapping $\alpha \rightarrow 1-\alpha ,$ so do the bound state
energies $E_{nl}.$ An important quantity of interest for the M-R potential
is the critical coupling constant $A_{c},$ which is that value of $A$ for
which the binding energy of the level in question becomes zero. Hence, using
Eq. (31), in atomic units $\hbar ^{2}=\mu =Z=e=1,$ we find the following
critical coupling constant

\begin{equation}
A_{c}=\left( n+1+\Lambda -\sqrt{l(l+1)d_{0}}\right) ^{2}-\Lambda (\Lambda
+1)+l(l+1)(1-d_{0}).
\end{equation}

Let us now find the corresponding radial part of the normalized wave
functions. Using $\sigma (z)$ and $\pi (z)$ in Eqs. (22) and (26), we obtain

\begin{equation}
\phi (z)=z^{\varepsilon ^{\prime }}(1-z)^{(\Lambda +1)/2},
\end{equation}
\begin{equation}
\rho (z)=z^{2\varepsilon ^{\prime }}(1-z)^{2\Lambda +1},
\end{equation}

\begin{equation}
y_{nl}(z)=C_{n}z^{-2\varepsilon ^{\prime }}(1-z)^{-(2\Lambda +1)}\frac{d^{n}%
}{dz^{n}}\left[ z^{n+2\varepsilon ^{\prime }}(1-z)^{n+2\Lambda +1}\right] .
\end{equation}
The functions $\ y_{nl}(z)$, up to a numerical factor, are in the form of\
Jacobi polynomials, i.e., $\ y_{nl}(z)\simeq P_{n}^{(2\varepsilon ^{\prime
},2\Lambda +1)}(1-2z),$ valid physically in the interval $(0\leq r<\infty $ $%
\rightarrow $ $0\leq z\leq 1)$ [30]. Therefore, the radial part of the wave
functions can be found by substituting Eqs. (33) and (35) into $%
R_{nl}(z)=\phi (z)y_{nl}(z)$ as

\begin{equation}
R_{nl}(z)=N_{nl}z^{\varepsilon ^{\prime }}(1-z)^{1+\Lambda
}P_{n}^{(2\varepsilon ^{\prime },2\Lambda +1)}(1-2z),
\end{equation}
where $\varepsilon $ and $\Lambda $ are given in Eqs. (24) and (30) and $%
N_{nl}$ is a normalization factor to be determined from the normalization
condition:

\begin{equation}
\int\limits_{0}^{\infty }\left| R_{nl}(r)\right|
^{2}dr=1=b\int\limits_{0}^{1}z^{-1}\left| R_{nl}(z)\right| ^{2}dz.
\end{equation}
This can be further written as

\begin{equation}
1=bN_{nl}^{2}\int\limits_{0}^{1}z^{2\varepsilon ^{\prime }-1}(1-z)^{2\Lambda
+2}\left[ P_{n}^{(2\varepsilon ^{\prime },2\Lambda +1)}(1-2z)\right] ^{2}dz.
\end{equation}
from which we obtain [26]

\begin{equation}
N_{nl}=\frac{1}{\sqrt{s(n)}},
\end{equation}
\[
s(n)=b(-1)^{n}\frac{\Gamma (n+2\Lambda +2)\Gamma (n+2\varepsilon ^{\prime
}+1)^{2}}{\Gamma (n+2\varepsilon ^{\prime }+2\Lambda +2)}
\]
\begin{equation}
\times \sum\limits_{p,r=0}^{n}\frac{(-1)^{p+r}\Gamma (n+2\varepsilon
^{\prime }+r-p+1)(p+2\Lambda +2)}{p!r!(n-p)!(n-r)!\Gamma (n+2\varepsilon
^{\prime }-p+1)\Gamma (2\varepsilon ^{\prime }+r+1)(n+2\varepsilon ^{\prime
}+r+2\Lambda +2)}.
\end{equation}

\section{Numerical Results}

To show the accuracy of the new approximation scheme, we have calculated the
energy eigenvalues for various $n$ and $l$ quantum numbers with two
different values of the parameters $\alpha .$ The results calculated by Eq.
(31) are compared with those obtained by a MATHEMATICA package programmed by
Lucha and Sch\"{o}berl [28] as shown in Table 1 for short-range (large $b$)
and long-range (small $b$) potentials. This is an illustration to assess the
validity and usefulness of our present calculation. The energy eigenvalues
for a few $HCl,CH,LiH$ and $CO$ diatomic molecules are presented in Tables 2
and 3. Lowest eigenvalues of $l=0,1,2,3$ are given at four values of $1/b$
in the range $0.025-0.1$ covering both weaker and stronger interaction to
demonstrate the generality of our results. The formalism is quite simple,
computationally efficient, reliable and illustrated very accurate.

\section{Discussions}

In this work, we have utilized N-U method and solved the radial ${\rm SE}$
for the M-R model potential with the angular momentum $l\neq 0$ states$.$ We
have derived the binding energy spectra in Eq. (31) and their corresponding
wave functions in Eq. (36).

Let us study special cases. We have shown that for $\alpha =0$ $(1)$, the
present solution reduces to the one of the Hulth\'{e}n potential [9,11]:

\begin{equation}
V^{(H)}(r)=-V_{0}\frac{e^{-\delta r}}{1-e^{-\delta r}},\text{ }%
V_{0}=Ze^{2}\delta ,\text{ }\delta =b^{-1}
\end{equation}
where $Ze^{2}$ is the strength and $\delta $ is the screening parameter and $%
b$ is the range of potential. If the potential is used for atoms, the $Z$ is
identified with the atomic number. This can be achieved by setting $\Lambda
=l,$ hence, the energy for $l\neq 0$ states

\begin{equation}
E_{nl}=-\frac{\left[ A-(n+l+1)^{2}\right] ^{2}\hbar ^{2}}{8\mu
b^{2}(n+l+1)^{2}}+\frac{\hbar ^{2}l(l+1)c_{0}}{2\mu b^{2}},\text{ \ }0\leq
n,l<\infty .
\end{equation}
and for $s$-wave ($l=0)$ states
\begin{equation}
E_{n}=-\frac{\left[ A-(n+1)^{2}\right] ^{2}\hbar ^{2}}{8\mu b^{2}(n+1)^{2}},%
\text{ \ }0\leq n<\infty
\end{equation}
Essentially, these results coincide with those obtained by the Feynman
integral method [17] and the standard way [18,19], respectively.
Furthermore, if taking $b=1/\delta $ and identifying $\frac{A\hbar ^{2}}{%
2\mu b^{2}}$ as $Ze^{2}\delta ,$ we are able to obtain
\begin{equation}
E_{nl}=-\frac{\mu \left( Ze^{2}\right) ^{2}}{2\hbar ^{2}}\left[ \frac{1}{%
n+l+1}-\frac{\hbar ^{2}\delta }{2Ze^{2}\mu }(n+l+1)\right] ^{2}+\frac{\hbar
^{2}l(l+1)c_{0}\delta ^{2}}{2\mu },
\end{equation}
and using the natural units $\hbar ^{2}=\mu =Z=e=1,$ we further obtain

\begin{equation}
E_{nl}=-\frac{1}{2}\left[ \frac{1}{n+l+1}-\frac{(n+l+1)}{2}\delta \right]
^{2}+\frac{l(l+1)c_{0}\delta ^{2}}{2}.
\end{equation}
The corresponding radial wave functions are expressed as

\begin{equation}
R_{nl}(r)=N_{nl}e^{-\delta \varepsilon ^{\prime }r}(1-e^{-\delta
r})^{l+1}P_{n}^{(2\varepsilon ^{\prime },2l+1)}(1-2e^{-\delta r}),
\end{equation}
where

\begin{equation}
\varepsilon ^{\prime }=\frac{\mu Ze^{2}}{\hbar ^{2}\delta }\left[ \frac{1}{%
n+l+1}-\frac{\hbar ^{2}\delta }{2Ze^{2}\mu }(n+l+1)\right] ,\text{ }0\leq
n,l<\infty ,
\end{equation}
which coincides for the ground state with G\"{o}n\"{u}l {\it et al.} [9] in
Eq. (6). In addition, for $\delta r\ll 1$ (i.e., $r/b\ll 1),$ the
Hulth\'{e}n potential turns to become a Coulomb potential: $V(r)=-Ze^{2}/r$
with energy levels and wavefunctions:

\[
E_{nl}=-\frac{\varepsilon _{0}}{(n+l+1)^{2}},\text{ }n=0,1,2,..
\]

\begin{equation}
\varepsilon _{0}=\frac{Z^{2}\hbar ^{2}}{2\mu a_{0}^{2}},\text{ }a_{0}=\frac{%
\hbar ^{2}}{\mu e^{2}}
\end{equation}
where $\varepsilon _{0}=13.6$ $eV$ and $a_{0}$ is Bohr radius for the
Hydrogen atom [3]. The wave functions are
\begin{equation}
R_{nl}=N_{nl}\exp \left[ -\frac{\mu Ze^{2}}{\hbar ^{2}}\frac{r}{\left(
n+l+1\right) }\right] r^{l+1}P_{n}^{\left( \frac{2\mu Ze^{2}}{\hbar
^{2}\delta (n+l+1)},2l+1\right) }(1+2\delta r)
\end{equation}
which coincide with Refs. [11,13].

\section{Cocluding Remarks}

In this work, we have used a new improved approximation to the centrifugal
term and determined approximately the arbitrary $l$-wave bound state
solution of the Schr\"{o}dinger equation with the M-R potential. The special
cases for $\alpha =0,1$ are discussed. The results are in good agreement
with those obtained by other methods for short potential range, small $%
\alpha $ and $l.$ We have also studied two special cases for $l=0,$ $l\neq 0$
and Hulth\'{e}n potential. The results we have ended up show that the N-U
method constitute a reliable alternative way in solving the exponential
potentials. The numerical results show that our results are in good
agreement with those obtained by using the MATHEMATICA program based on the
numerical integration procedure [28]. This means that the approximation in
Eq. (16) is a good approximation since the energy are very close to the
onesobtained in [28]. Furthermore, we have applied this approximation in
obtaining the energy bound states ($-E_{nl}$) for a few $HCl,$ $CH,$ $LiH$
and $CO$ diatomic molecules.

\acknowledgments
This research is partially supported by the Scientific and Technological
Research Council of Turkey.

\bigskip

\bigskip \baselineskip= 2\baselineskip
\bigskip

\begin{table}[tbp]
\caption{Bound state energy eigenvalues ($-E_{nl}$) (in atomic units) for
the Manning-Rosen potential as a function of $1/b$ for $%
2p,3p,3d,4p,4d,4f,5p,5d,5f,5g,6p,6d,6f$ and $6g$ states with $\protect\alpha %
=0.75,$ $\protect\alpha =1.5$ and $A=2b.$}
\begin{tabular}{llllllll}
&  & $\alpha =0.75$ &  &  & $\alpha =1.5$ &  &  \\
states & $1/b$ & present & previous [26] & Lucha et al [28] & present &
previous [26] & Lucha et al [28] \\
\tableline$2p$ & $0.025$ & $0.1205297$ & $0.1205793$ & $0.1205271$ & $%
0.0899732$ & $0.0900229$ & $0.0899708$ \\
& $0.050$ & $0.1082245$ & $0.1084228$ & $0.1082151$ & $0.0800489$ & $%
0.0802472$ & $0.0800400$ \\
& $0.075$ & $0.0964658$ & $0.0969120$ & $0.0964469$ & $0.0705870$ & $%
0.0710332$ & $0.0705701$ \\
& $0.100$ & $0.0852807$ & $0.0860740$ &  & $0.0569224$ & $0.0577157$ &  \\
$3p$ & $0.025$ & $0.0458800$ & $0.0459297$ & $0.0458779$ & $0.0369154$ & $%
0.0369651$ & $0.0369134$ \\
& $0.050$ & $0.0350689$ & $0.0352672$ & $0.0350633$ & $0.0272736$ & $%
0.0274719$ & $0.0272696$ \\
& $0.075$ & $0.0255647$ & $0.0260110$ & $0.0255654$ & $0.0189388$ & $%
0.0193850$ & $0.0189474$ \\
& $0.100$ & $0.0173676$ & $0.0181609$ &  & $0.0119110$ & $0.0127043$ &  \\
$3d$ & $0.025$ & $0.0447812$ & $0.0449299$ & $0.0447743$ & $0.0394857$ & $%
0.0396345$ & $0.0394789$ \\
& $0.050$ & $0.0337133$ & $0.0343082$ & $0.0336930$ & $0.0294680$ & $%
0.0300629$ & $0.0294496$ \\
& $0.075$ & $0.0237782$ & $0.0251168$ & $0.0237621$ & $0.0204734$ & $%
0.0218121$ & $0.0204663$ \\
$4p$ & $0.025$ & $0.0208112$ & $0.0208608$ & $0.0208097$ & $0.0171753$ & $%
0.0172249$ & $0.0171740$ \\
& $0.050$ & $0.0117308$ & $0.0119292$ & $0.0117365$ & $0.0089036$ & $%
0.0091019$ & $0.0089134$ \\
& $0.075$ & $0.0050311$ & $0.0054773$ & $0.0050945$ & $0.0031016$ & $%
0.0035478$ & $0.0031884$ \\
$4d$ & $0.025$ & $0.0203068$ & $0.0204555$ & $0.0203017$ & $0.0182162$ & $%
0.0183649$ & $0.0182115$ \\
& $0.050$ & $0.0109792$ & $0.0115742$ & $0.0109904$ & $0.0094998$ & $%
0.0100947$ & $0.0095167$ \\
& $0.075$ & $0.0038661$ & $0.0052047$ & $0.0040331$ & $0.0029422$ & $%
0.0042808$ & $0.0031399$ \\
$4f$ & $0.025$ & $0.0199911$ & $0.0202887$ & $0.0199797$ & $0.0186247$ & $%
0.0189223$ & $0.0186137$ \\
& $0.050$ & $0.0102384$ & $0.0114284$ & $0.0102393$ & $0.0093953$ & $%
0.0105852$ & $0.0094015$ \\
& $0.075$ & $0.0024162$ & $0.0050935$ & $0.0026443$ & $0.0019754$ & $%
0.0046527$ & $0.0022307$ \\
$5p$ & $0.025$ & $0.0098080$ & $0.0098576$ & $0.0098079$ & $0.0080812$ & $%
0.0081308$ & $0.0080816$ \\
$5d$ & $0.025$ & $0.0095150$ & $0.0096637$ & $0.0095141$ & $0.0085415$ & $%
0.0086902$ & $0.0085415$ \\
$5f$ & $0.025$ & $0.0092862$ & $0.0095837$ & $0.0092825$ & $0.0086647$ & $%
0.0089622$ & $0.0086619$ \\
$5g$ & $0.025$ & $0.0090440$ & $0.0095398$ & $0.0090330$ & $0.0086252$ & $%
0.0091210$ & $0.0086150$ \\
$6p$ & $0.025$ & $0.0043555$ & $0.0044051$ & $0.0043583$ & $0.0034838$ & $%
0.0035334$ & $0.0034876$ \\
$6d$ & $0.025$ & $0.0041574$ & $0.0043061$ & $0.0041650$ & $0.0036722$ & $%
0.0038209$ & $0.0036813$ \\
$6f$ & $0.025$ & $0.0039677$ & $0.0042652$ & $0.0039803$ & $0.0036631$ & $%
0.0039606$ & $0.0036774$ \\
$6g$ & $0.025$ & $0.0037470$ & $0.0042428$ & $0.0037611$ & $0.0035464$ & $%
0.0040422$ & $0.0035623$%
\end{tabular}
\end{table}

\begin{table}[tbp]
\caption{Bound state energy eigenvalues ($-E_{nl}$) (in $eV$) for $HCl$ and $%
CH$ for $2p,3p,3d,4p,4d,4f,5p,5d,5f,5g,6p,6d,6f$and $6g$ states with $\hbar
c=1973.29$ $eV$ $A^{\circ },$ $\protect\mu _{HCl}=0.9801045$ $amu,$ $\protect%
\mu _{CH}=0.929931$ $amu$ and $A=2b.$}
\begin{tabular}{llllllll}
states & $1/b$\tablenotetext[1]{$b$ is in $pm$.}\tablenotemark[1] & $HCl/$ $%
\alpha =0,1$ & $\alpha =0.75$ & $\alpha =1.5$ & $CH/$ $\alpha =0,1$ & $%
\alpha =0.75$ & $\alpha =1.5$ \\
\tableline$2p$ & $0.025$ & $4.80941188$ & $5.14067096$ & $3.83741636$ & $%
5.06889891$ & $5.41803073$ & $4.04446034$ \\
& $0.050$ & $4.30992001$ & $4.61584459$ & $3.41413694$ & $4.54245745$ & $%
4.86488789$ & $3.59834329$ \\
& $0.075$ & $3.83285565$ & $4.11432861$ & $3.01058097$ & $4.03965355$ & $%
4.33631311$ & $3.17301386$ \\
& $0.100$ & $3.37821878$ & $3.63612726$ & $2.42777890$ & $3.56048721$ & $%
3.83231089$ & $2.55876729$ \\
$3p$ & $0.025$ & $1.86422242$ & $1.95681272$ & $1.57446670$ & $1.96480468$ &
$2.06239060$ & $1.65941548$ \\
& $0.050$ & $1.41471071$ & $1.49571070$ & $1.16323608$ & $1.49104002$ & $%
1.57641028$ & $1.22599733$ \\
& $0.075$ & $1.02094947$ & $1.09035060$ & $0.80775166$ & $1.07603378$ & $%
1.14917938$ & $0.85133310$ \\
& $0.100$ & $0.68293440$ & $0.74074096$ & $0.50801342$ & $0.71978146$ & $%
0.78070691$ & $0.53542279$ \\
$3d$ & $0.025$ & $1.85999327$ & $1.90994571$ & $1.68408920$ & $1.96034735$ &
$2.01299493$ & $1.77495255$ \\
& $0.050$ & $1.39779410$ & $1.43789211$ & $1.25682731$ & $1.47321069$ & $%
1.51547215$ & $1.32463817$ \\
& $0.075$ & $0.98288709$ & $1.01415428$ & $0.87320241$ & $1.03591778$ & $%
1.06887196$ & $0.92031517$ \\
& $0.100$ & $0.61526795$ & $0.63872794$ & $0.53322303$ & $0.64846412$ & $%
0.67318987$ & $0.56199254$ \\
$4p$ & $0.025$ & $0.85089842$ & $0.88761210$ & $0.73253860$ & $0.89680780$ &
$0.93550233$ & $0.77206199$ \\
& $0.050$ & $0.47136150$ & $0.50032556$ & $0.37974364$ & $0.49679334$ & $%
0.52732013$ & $0.40023233$ \\
& $0.075$ & $0.19422206$ & $0.21457922$ & $0.13228479$ & $0.20470112$ & $%
0.22615662$ & $0.13942208$ \\
$4d$ & $0.025$ & $0.84666927$ & $0.86609664$ & $0.77693119$ & $0.89235047$ &
$0.91282602$ & $0.81884974$ \\
& $0.050$ & $0.45444489$ & $0.46826797$ & $0.40517060$ & $0.47896401$ & $%
0.49353290$ & $0.42703117$ \\
& $0.075$ & $0.15615968$ & $0.16489027$ & $0.12548533$ & $0.16458512$ & $%
0.17378676$ & $0.13225577$ \\
$4f$ & $0.025$ & $0.84032554$ & $0.85263452$ & $0.79435667$ & $0.88566447$ &
$0.89863756$ & $0.83721539$ \\
& $0.050$ & $0.42906997$ & $0.43667458$ & $0.40071582$ & $0.45222001$ & $%
0.46023492$ & $0.42233604$ \\
& $0.075$ & $0.09906611$ & $0.10305395$ & $0.08425354$ & $0.10441112$ & $%
0.10861411$ & $0.08879935$ \\
$5p$ & $0.025$ & $0.40106735$ & $0.41831847$ & $0.34466933$ & $0.42270654$ &
$0.44088842$ & $0.36326562$ \\
$5d$ & $0.025$ & $0.39683820$ & $0.40581936$ & $0.36429895$ & $0.41824921$ &
$0.42771494$ & $0.38395434$ \\
$5f$ & $0.025$ & $0.39049447$ & $0.39606358$ & $0.36955620$ & $0.41156321$ &
$0.41743279$ & $0.38949523$ \\
$5g$ & $0.025$ & $0.38203616$ & $0.38573290$ & $0.36787081$ & $0.40264543$ &
$0.40654473$ & $0.38771891$ \\
$6p$ & $0.025$ & $0.17707786$ & $0.18576580$ & $0.14858723$ & $0.18663192$ &
$0.19578861$ & $0.15660410$ \\
$6d$ & $0.025$ & $0.17284871$ & $0.17731423$ & $0.15662014$ & $0.18217459$ &
$0.18688105$ & $0.16507042$ \\
$6f$ & $0.025$ & $0.16650498$ & $0.16922609$ & $0.15623470$ & $0.17548859$ &
$0.17835652$ & $0.16466420$ \\
$6g$ & $0.025$ & $0.15804667$ & $0.15981241$ & $0.15125669$ & $0.16657392$ &
$0.16843493$ & $0.15941759$%
\end{tabular}
\end{table}

\begin{table}[tbp]
\caption{Bound state energy eigenvalues $(-E_{nl})$ (in $eV$) for $LiH$ and $%
CO$ for $2p,3p,3d,4p,4d,4f,5p,5d,5f,5g,6p,6d,6f$and $6g$ states with $\hbar
c=1973.29$ $eV$ $A^{\circ },$ $\protect\mu _{LiH}=0.8801221$ $amu,$ $\protect%
\mu _{CO}=6.8606719$ $amu$ and $A=2b.$}
\begin{tabular}{llllllll}
states & $1/b$\tablenotemark[1]\tablenotetext[1]{$b$ is in $pm$.} & $LiH/$ $%
\alpha =0,1$ & $\alpha =0.75$ & $\alpha =1.5$ & $CO/$ $\alpha =0,1$ & $%
\alpha =0.75$ & $\alpha =1.5$ \\
\tableline$2p$ & $0.025$ & $5.35576397$ & $5.72465427$ & $4.27334918$ & $%
1.37443170$ & $0.73438794$ & $0.54852071$ \\
& $0.050$ & $4.79952952$ & $5.14020732$ & $3.80198495$ & $1.23262476$ & $%
0.65941210$ & $0.48773809$ \\
& $0.075$ & $4.26827035$ & $4.58171881$ & $3.35258477$ & $1.09782989$ & $%
0.58776634$ & $0.43008673$ \\
& $0.100$ & $3.76198647$ & $4.04919351$ & $2.70357604$ & $0.97004711$ & $%
0.51945126$ & $0.34682857$ \\
$3p$ & $0.025$ & $2.07599922$ & $2.17910783$ & $1.75332707$ & $0.53294169$ &
$0.27954710$ & $0.22492577$ \\
& $0.050$ & $1.57542270$ & $1.66562433$ & $1.29538040$ & $0.40541491$ & $%
0.21367481$ & $0.16617803$ \\
& $0.075$ & $1.13692993$ & $1.21421508$ & $0.89951273$ & $0.29442115$ & $%
0.15576572$ & $0.11539410$ \\
& $0.100$ & $0.76051617$ & $0.82488959$ & $0.56572406$ & $0.19995917$ & $%
0.10582106$ & $0.07257398$ \\
$3d$ & $0.025$ & $2.07128963$ & $2.12691670$ & $1.87540274$ & $0.53233752$ &
$0.27285176$ & $0.24058626$ \\
& $0.050$ & $1.55658435$ & $1.60123752$ & $1.39960364$ & $0.40299823$ & $%
0.20541494$ & $0.17954832$ \\
& $0.075$ & $1.09454364$ & $1.12936281$ & $0.97239872$ & $0.29098362$ & $%
0.14488044$ & $0.12474428$ \\
& $0.100$ & $0.68516276$ & $0.71128782$ & $0.59379748$ & $0.19029245$ & $%
0.09124764$ & $0.07617538$ \\
$4p$ & $0.025$ & $0.94756010$ & $0.98844538$ & $0.81575544$ & $0.24341803$ &
$0.12680283$ & $0.10464928$ \\
& $0.050$ & $0.52490845$ & $0.55716284$ & $0.42288275$ & $0.13588423$ & $%
0.07147570$ & $0.05424956$ \\
& $0.075$ & $0.21628580$ & $0.23895554$ & $0.14731241$ & $0.05821126$ & $%
0.03065444$ & $0.01889799$ \\
$4d$ & $0.025$ & $0.94285141$ & $0.96448574$ & $0.86519105$ & $0.24281386$ &
$0.12372917$ & $0.11099113$ \\
& $0.050$ & $0.50607010$ & $0.52146349$ & $0.45119822$ & $0.13346755$ & $%
0.06658523$ & $0.05788202$ \\
& $0.075$ & $0.17389951$ & $0.18362190$ & $0.13974054$ & $0.05277373$ & $%
0.02355596$ & $0.01792663$ \\
$4f$ & $0.025$ & $0.93578703$ & $0.94949432$ & $0.88459607$ & $0.24190761$ &
$0.12180599$ & $0.11348051$ \\
& $0.050$ & $0.47781258$ & $0.48628108$ & $0.44623738$ & $0.12984253$ & $%
0.06238263$ & $0.05724561$ \\
& $0.075$ & $0.11032008$ & $0.11476093$ & $0.09382479$ & $0.04461744$ & $%
0.01472212$ & $0.01203632$ \\
$5p$ & $0.025$ & $0.44662885$ & $0.46583971$ & $0.38382398$ & $0.11489375$ &
$0.05976030$ & $0.04923890$ \\
$5d$ & $0.025$ & $0.44191926$ & $0.45192068$ & $0.40568353$ & $0.11428958$ &
$0.05797470$ & $0.05204316$ \\
$5f$ & $0.025$ & $0.43485488$ & $0.44105664$ & $0.41153801$ & $0.11338333$ &
$0.05658100$ & $0.05279420$ \\
$5g$ & $0.025$ & $0.42543570$ & $0.42955239$ & $0.40966116$ & $0.11217500$ &
$0.05510518$ & $0.05255343$ \\
$6p$ & $0.025$ & $0.19719402$ & $0.20686891$ & $0.16546683$ & $0.05089620$ &
$0.02653820$ & $0.02122693$ \\
$6d$ & $0.025$ & $0.19248443$ & $0.19745724$ & $0.17441228$ & $0.05029203$ &
$0.02533082$ & $0.02237450$ \\
$6f$ & $0.025$ & $0.18542005$ & $0.18845028$ & $0.17398306$ & $0.04938577$ &
$0.02417537$ & $0.02231944$ \\
$6g$ & $0.025$ & $0.17600087$ & $0.17796720$ & $0.16843954$ & $0.04817743$ &
$0.02283054$ & $0.02160829$%
\end{tabular}
\end{table}

\end{document}